\documentstyle[twocolumn,aps,epsfig]{revtex}
\begin{document}
\twocolumn[
\hsize\textwidth\columnwidth\hsize\csname @twocolumnfalse\endcsname

\title{ Two-frequency shell-model calculations for $p$-shell nuclei }
\author{L. Coraggio,$^1$ A. Covello,$^1$ A. Gargano,$^1$ N. Itaco,$^1$
and T. T. S. Kuo$^2$}
\address{$^1$Dipartimento di Scienze Fisiche, Universit\`a
di Napoli Federico II, and Istituto Nazionale di Fisica Nucleare \\
Complesso Universitario di Monte S. Angelo, Via Cintia, I-80126 Napoli,
Italy\\
$^2$Department of Physics, State University of New York at Stony Brook,
Stony Brook, New York 11794}
\date{\today}
\maketitle
\begin{abstract}
\begin{small}

We have studied $p$-shell
nuclei  using a two-frequency shell-model approach
with an effective interaction derived from
the Bonn-A nucleon-nucleon potential by means of a  $G$-matrix
folded-diagram method.
Shell-model wave functions
of two different oscillator constants , $\hbar \omega_{\rm in}$ and $\hbar
\omega_{\rm out}$, are employed,
one for the  inner $0s$ core orbit and the other for the outer valence
orbits, respectively.
The binding energies, energy spectra, and electromagnetic properties are
calculated and compared with experiment.
A quite satisfactory agreement with the experimental data is obtained, which
is in some cases even better
than that produced by large-basis shell-model calculations.

\end{small}
\end{abstract}
\draft
\vspace{4mm}

]

\section{Introduction}

The $p$-shell nuclei have long been the subject of theoretical interest. The
first shell-model study of these nuclei was
performed by Cohen and Kurath in  1965 \cite{cohen65}. In this work, which
has been a point of reference for later studies,
a successful description of the $p$-shell nuclei was given  by taking
$^{4}$He as a closed core and
letting the valence nucleons occupy the $0p$ shell. The fifteen matrix
elements of the two-body interaction and
the two single-particle energies were determined by making a least-squares
fit to selected observed energy levels.
>From  then  on, several other shell-model calculations have been  performed
in the $0p$
model space employing different kinds of effective interactions [2-5]. It
should  be mentioned that the calculations of
Ref. \cite{halb66} represent the first attempt to use a realistic effective
interaction
for these light nuclei.

In recent years, the study of $p$-shell nuclei has become a subject of
special interest
owing to the discovery of new aspects of their structure. One main result
has been the observation
for some neutron-rich nuclei, such as $^{6}$He and
$^{11}$Li, of abnormally large interaction and reaction cross-sections
\cite{tan85,tan85b}. These nuclei have a very small one- and
two-neutron separation energy and have been described as having a halo
structure \cite{han87}, namely an
extended neutron distribution surrounding a tightly bound inner core.

During the last decade substantial progress in computational techniques has
set the stage
for more ambitious calculations of the structure of light nuclei.
As regards shell-model studies, calculations in a $(0+2)\hbar \omega$ model
space have been
performed in the early 1990s [9-11]. More recently, larger multi-$\hbar
\omega$ spaces have been used \cite{kara97,nav98}.
In particular, large-basis no-core calculations have been carried out
\cite{nav98}
making use of an effective interaction derived from the Reid 93
nucleon-nucleon ($NN$) potential. Alternatively,
there has been a variety of studies in terms of clusters (see  Ref.
\cite{kuku95} for a comprehensive list of references).
In this context, three-body model approaches aimed at describing the
structure of halo nuclei have been
developed \cite{kuku95,zhuc00}.

To end this brief review of the various approaches to the study of $p$-shell
nuclei, the
quantum Monte Carlo calculations of Refs. \cite{pud97,wiri00} should be
mentioned.
Within this approach properties of nuclei with $A \leq 8$ are calculated
directly from
bare two-nucleon and three-nucleon forces.

A new approach, the two-frequency shell-model  (TFSM), has been recently
proposed in Refs.  \cite{kuo96tf,kuo97}.
Within the TFSM, the model space effective interaction $V_{\rm eff}$
is derived from the free $NN$ potential by way of a $G$-matrix folded
diagram method. Its peculiar feature consists in calculating
the $G$ matrix in a space composed of harmonic oscillator wave functions
with two different
oscillator constants, $\hbar \omega_{\rm in}$ and  $\hbar \omega_{\rm out}$,
for the core and the valence orbits, respectively
(the length parameters $b_{\rm in}$ and $b_{\rm out}$ will be also used from
now on, with $b=(\hbar/m \omega)^{1/2}$).
Note that $b_{\rm out}$ is chosen substantially larger than $b_{\rm in}$.
This idea reflects the fact that the valence
nucleons in $p$-shell nuclei are
spatially more extended than those of the core. Actually, these nuclei may
be thought of as a $^{4}$He nucleus with
loosely attached outer nucleons. This feature may be taken into account by
including several major shells in the
ordinary one-frequency shell model. We shall see that the TFSM, allowing
different length parameters for
the valence and core orbits, provides a simple and effective alternative.
It may be interesting to mention that, albeit in quite different contexts,
this idea was also considered
in two earlier works
\cite{elton61,bouten89}.

The outline of the paper is as follows. In Sec. I the derivation of the
effective interaction
from a realistic $NN$ potential is described. Our results are presented and
compared with the
experimental data in Sec. III. Sec. IV contains a summary of our
conclusions.

\section{Derivation of the  effective interaction}

Here, we describe how to derive the effective interaction
from a realistic $NN$ potential $V_{NN}$ within the framework of the TFSM.

As usual, one  starts from a nuclear many-body problem of the form
$H\Psi_{\mu}=E_{\mu}\Psi_{\mu}$ with $H=T+V_{NN}$, where $T$ denotes
the kinetic energy.
This many-body problem can be formally reduced \cite{kuos} to a model space
(usually referred to as the $P$-space) problem of the form

\begin{equation}
H_{\rm eff}P\Psi_{\mu}=E_{\mu}P\Psi_{\mu};~H_{\rm eff}=H_0+V_{\rm eff},
\end{equation}
where the eigenvalues $E_{\mu}$ are a subset of the eigenvalues of the
original Hamiltonian in the full space, $\mu= 1,2 \cdots d$, with $d$
denoting the dimension of the $P$ space.
In Eq. (1)  $V_{\rm eff}$ is the model-space effective interaction
and $H_0=T+U$ the unperturbed Hamiltonian, $U$ being an auxiliary potential
introduced to define a convenient single-particle (sp) basis.
This is  chosen to be a harmonic oscillator potential. Note that our $P$
space is defined
in terms of the eigenfunctions of $H_0$.

The model-space effective interaction
$V_{\rm eff}$ may be written  \cite{kuos} as a folded-diagram series

\begin{equation}
V_{\rm eff}= \hat{Q} - \hat{Q}^{'}\int\hat{Q}
+ \hat{Q}^{'}\int\hat{Q}\int\hat{Q} - \hat{Q}^{'}\int\hat{Q}\int\hat{Q}
\int\hat{Q}\cdots,
\end{equation}
where $\int$ denotes a generalized fold, and $\hat{Q}^{'}$ and $\hat{Q}$
represent the $\hat Q$-box, composed of irreducible valence-linked diagrams.
$\hat Q'$ is obtained from  $\hat Q$ by removing first-order diagrams.
Because of the strong repulsive core contained in all modern $NN$
potentials,
as a first step we need to derive the model-space $G$-matrix corresponding
to the chosen $V_{NN}$, and then calculate the $\hat Q$-box from irreducible
diagrams with $G$-matrix vertices. The  Brueckner $G$-matrix
is defined by the integral equation \cite{kkko,muet92}

\begin{equation}
G(\omega)=V+V Q_2\frac{1}
{\omega-Q_2TQ_2}Q_2G(\omega),
\end{equation}
where $\omega$ is an energy variable, $T$ is the two-nucleon kinetic energy
and $V$ represents the $NN$ potential.
$Q_2$ is a two-body Pauli exclusion
operator, whose complement $P_{2}=1-Q_{2}$ defines the space within which
the $G$ matrix is calculated. The role of $Q_{2}$ in Eq. (3) is to prevent
double counting, namely the intermediate
states allowed for $G$ must be outside of the $P_{2}$ space.
Note that our $G$ matrix has orthogonalized plane-wave functions as
intermediate states
while the operator $Q_{2}$ is defined in terms of harmonic
oscillator wave functions as

\begin{equation}
Q_{2}= \sum_{{\rm all} \ a b} Q(ab) |ab \rangle \langle ab|,
\end{equation}
where $Q(ab)=0$, if $b \leq n_{1}$, $a \leq n_{3}$, or $b \leq n_{2}$,
$a \leq n_{2}$, or $b \leq n_{3}$, $a \leq n_{1}$, and $Q(ab)=1$
otherwise.
The boundary of  $Q_{2}$ is specified
by the three numbers $n_{1}$,
$n_{2}$, and $n_{3}$, each representing a sp orbit
(the orbits are numbered starting from the bottom of the oscillator well).
In particular, $n_{1}$ is the number of orbits below the
Fermi surface of the doubly magic core, $n_{2}$  fixes the orbit
above which  the passive sp states start, and $n_{3}$ denotes
the limit of the $P_{2}$ space.

It should be noted that in the calculation of $G$ the space of active
sp states,
i.e. the levels
between $n_{1}$ and $n_{2}$, may be different from the model space within
which $V_{\rm eff}$ is defined. Several arguments for choosing
the former larger
than the latter are given in Ref. \cite{kkko}. Generally, $n_{2}$ is fixed
so as to include two major shells above the Fermi surface. In this paper,
we consider the $p$-shell nuclei with $^{4}$He as a core, thus we have
$n_{1}=1$. Then we take $n_{2}=6$ so as to include all the five orbits of
the $p$ and $sd$ shells above the Fermi surface. As regards
$n_{3}$, it should be infinite, but in practice it is chosen to be
a large but finite number. Namely, calculations are performed for
increasing values of $n_{3}$ until numerical results become stable.
For the present case, we have found that a choice of $n_{3}=21$ turns
out to be quite adequate.

>From the above it is clear that the reaction matrix $G$  depends on the
space $P_{2}$ and will be different for different choices
of this  space. In the TFSM approach the $P_{2}$ space is defined in terms
of harmonic oscillator wave functions with two different length parameters
$b_{\rm in}$ and $b_{\rm out}$, the former for the inner core orbits and the
latter for the outer valence orbits.
As already discussed in the Introduction, this choice, with
$b_{\rm out}$ larger than $b_{\rm in}$, allows us to give an appropriate
description of the $p$-shell nuclei.

The presence of the Pauli operator $Q_2$ adds considerable difficulty
to the calculation of the above $G$-matrix. However, an accurate treatment
of it can be carried out using a matrix inversion
method \cite{kkko,tsaikuo}. With this method, the
exact solution  of the $G$-matrix equation (3) reads
\begin{equation}
G=G_F+\Delta G,
\end{equation}
where the ``free" $G$ matrix is
\begin{equation}
G_{F}(\omega)=V+V\frac{1}{\omega-T}G_{F}(\omega),
\end{equation}
and  the Pauli correction term ${\Delta G}$ is given by
\begin{equation}
\Delta G(\omega)=-G_F(\omega)\frac{1}{e}P_2\frac{1}
{P_2[1/e+(1/e)G_F(\omega)(1/e)]P_2}
P_2\frac{1}{e}G_F(\omega),
\end{equation}
where $e=\omega-T$.

The central ingredient for calculating the above $G$ matrix are the matrix
 elements
of $G_F$ within the $P_{2}$ space. As there is no Pauli
projection operator for $G_F$, the calculation of its momentum space
($k$-space) matrix elements is relatively easy
and has been carried out using the standard momentum-space matrix inversion
method \cite{kkko}. Similarly we have calculated the $k$-space
matrix elements of
$1/e~G_F$, $G_F~1/e$ and $1/e~G_F~1/e$. For shell model calculations,
however, we need
the matrix elements of these operators between  oscillator basis wave
functions.
In our two-frequency approach, sp wave functions of two different length
parameters are employed, i.e. our basis consists of
both $\phi^{\rm in}_{n}$ and $\phi^{\rm out}_{n}$, the oscillator
wave functions
with length parameters $b_{\rm in}$ and $b_{\rm out}$, respectively.
As a consequence, we also have to calculate matrix elements
such as $\langle \phi^{\rm in}_{1} ~\phi^{\rm out}_{2} \vert G_F
\vert \phi^{\rm out}_{3} ~\phi^{\rm out}_{4} \rangle $.

To calculate matrix elements of the above type, a standard procedure
is to first transform the wave functions to the
RCM (relative and center of mass) representation.
For the above matrix element, the two sp wave functions
in the ket $\vert \phi^{\rm out}_{3} ~\phi^{\rm out}_{4} \rangle$
have identical length
parameters. While the RCM transformation for this state
can be easily carried out using the well-known Moshinsky transformation
brackets,
this is more complicated for the bra
$\langle \phi^{\rm in}_{1} ~\phi^{\rm out}_{2} \vert$,
as the two sp wave functions have different length parameters.
We have overcome this difficulty by expanding $\phi^{\rm in}$ in terms of
$\phi^{\rm out}$ or $vice~ versa$. By way of illustration, for the above
case
we have expanded the bra
$\langle \phi^{\rm in}_{1} ~\phi^{\rm out}_{2} \vert$ as

\begin{equation}
 \langle \phi^{\rm in}_{1} ~\phi^{\rm out}_{2} \vert = \sum _{n=0,N}
 C_{1,n} \langle \phi^{\rm out}_{n} ~\phi^{\rm out}_{2} \vert.
\end{equation}
With this expansion, the above matrix element becomes a linear combination
of $\langle \phi^{\rm out}_{n} ~\phi^{\rm out}_{2} \vert G_F
\vert \phi^{\rm out}_{3} ~\phi^{\rm out}_{4} \rangle $,
which is a one-frequency matrix element and can be readily evaluated.
 We have found
that this expansion can be carried out quite accurately by including
only a small number of terms, typically
$N\leq10$, in Eq. (8).
Similarly, we have calculated the mixed-frequency matrix
elements of $1/e~G_F$, $G_F~1/e$ and $1/e~G_F~1/e$.
In this way  the $G$ matrix of Eq. (5) is finally obtained.

A problem inherent in the TFSM may be mentioned. We must require the sp wave
functions $\phi_{n}$ to form an orthonormal basis.
This requirement is usually not satisfied by wave
functions of different length parameters. For instance,
$\phi^{\rm in}_{0p_{3/2}}$ is not orthogonal to $\phi^{\rm out}_{1p_{3/2}}$
when $b_{\rm in}$ is not equal to $b_{\rm out}$. In the present
work we  consider nuclei with several
nucleons in the orbits $0p_{3/2}$ and $0p_{1/2}$
outside the $^{4}$He  core. We have used a short length
parameter $b_{\rm in}$ for the $0s_{1/2}$ orbit
and a long length parameter $b_{\rm out}$ for the orbits mentioned
above. In  the calculation of the Pauli correction terms for
the $G$-matrix and in the derivation of $V_{\rm eff}$,
some higher orbits, such as the $1s_{1/2}$ orbit, are also needed.
To ensure their orthogonality with the core orbit,
we have also used $b_{\rm in}$ for the $1s_{1/2}$ and higher $s$
orbits ($b_{\rm out}$ is used for all the other higher orbits).
We shall further discuss this point later.

Using the above $G$ matrix, we can now calculate the  $\hat Q$-box of Eq.
(2).
This is done by including the seven first- and second-order irreducibile
valenced-linked $G$-matrix diagrams \cite{jiang92,hjorth95}, as shown in
Fig. 1.
 After the
$\hat Q$-box is calculated, $V_{\rm eff}$ is obtained by summing up the
folded-diagram series (2) to all orders by means of the Lee-Suzuki iteration
method \cite{lesu80,sule80}. This last step can be performed in an
essentially exact way for a given $\hat Q$-box.
Note that the $G$ matrix is energy
dependent in that it depends on the starting energy $\omega$.
The folded-diagram effective interaction given by Eq. (2) is, however,
energy independent \cite{kuos}.

Before closing this section we should remark that in our derivation of
$V_{\rm eff}$ only the calculation of the $\hat Q$-box requires certain
approximations. In fact, we have neglected
its $G$-matrix diagrams beyond the second-order
ones.
In Refs. \cite{hjorth95} and \cite{hjorth96} the role of third-order
diagrams was investigated
within the
framework of standard shell-model calculations. It was shown that
for the $sd$ nuclei
the third-order contributions produce  a change of about $10-15\%$ in
the effective interaction, which  reduces to only $5\%$ or less
for heavier nuclei (in this case only the $T=1$ matrix elements were
investigated).
In the TFSM approach one
expects these higher-order diagrams to be  even
smaller. In fact, the contribution  of the D7 diagram of Fig. 1, which is a
second-order core-polarization diagram and contributes  a significant
correction to the $G$ matrix,
is rather small  when the length parameter $b_{\rm out}$ becomes
significantly larger than  $b_{\rm in}$. Diagonal matrix elements
of this diagram  for the states
$\vert (p_{3/2})^{2};T=1,J=0 \rangle $ and
$\vert p_{3/2}p_{1/2};T=0,J=1 \rangle $ are shown in Fig. 2 as a function of
the outer length parameter $b_{\rm out}$. This parameter ranges from
1.45 to 2.50 fm while $b_{\rm in}$ is kept fixed at 1.45 fm. The Bonn-A
realistic $NN$ potential \cite{mach89} is used. We see that the
diagram D7 is already largely suppressed when $b_{\rm out}$
becomes nearly 2.0 fm.
We have also calculated several typical third-order diagrams and have
found that their contribution to the matrix elements of $V_{\rm eff}$
decreases  by an order of magnitude as $b_{\rm out}$ goes from 1.45 to 2.0
fm.
This is a consequence of the fact that
increasing $b_{\rm out}$ corresponds to  increasing the average distance
between the core and valence nucleons, thus reducing the overlap
between their wave functions.

We should like to recall that to ensure the orthogonality we have used the
same length parameter
$b_{\rm in}$ for not only the $0s_{1/2}$ but also the $1s_{1/2}$ and
other $s$ orbits. Using $b_{\rm in}$ only
for the core $0s_{1/2}$ orbit and  $b_{\rm out}$ for the other
$s$ orbits would of course require an  orthogonalization procedure,
which is numerically
more involved than our present treatment. We are currently examining
this point.

\section{Results and comparison with experiment}

Within the framework of the TFSM we have carried out calculations for the
$p$-shell nuclei with $A\leq9$.
Results of this study for $A=8$ nuclei have already been presented in
\cite{kuo99,gar00},
together with those obtained in a standard one-frequency
shell-model calculation. In these papers
comparison between one- and two-frequency calculations has evidenced the
merit of
the latter approach with respect to the former.

We have assumed that the
doubly magic $^{4}$He is a closed core and let the valence particles
occupy the two orbits $0p_{3/2}$ and $0p_{1/2}$. As regards the  sp spacing
between these
two levels, we have taken it from the experimental spectrum \cite{nndc} of
$^{5}$He,
namely $\epsilon_{1/2}-\epsilon_{3/2}=4.0$ MeV, while we have fixed the sp
energy
$\epsilon_{3/2}$ at 0.886 MeV, which is the experimental one-neutron
separation
energy for $^{5}$He \cite{audi95}.
It should be noted that the excitation energy of the first $\frac{1}{2}^{-}$
state in $^{5}$He, which is a very broad resonance, has a large error bar
($\pm1$ MeV).
The effective interaction has been derived
from the Bonn-A free $NN$ potential, as described
in Sec. II. All results presented in this paper have been obtained by using
the OXBASH
shell model code \cite{oxb}.

The $b_{\rm in}$ parameter used for the $0s_{1/2}$ orbit was
fixed at 1.45 fm \cite{kuo96tf}, while $b_{\rm out}$  was allowed to vary
from 1.45 to 2.50 fm.
In Table I we report  the experimental ground-state
binding energies \cite{audi95} for nuclei with $6 \leq A \leq 9$ and compare
them with the calculated ones for
$b_{\rm out}=1.45$, 1.75, 2.00, 2.25, and 2.50 fm.
The theoretical values have been obtained by adding to our calculated
ground-state energies the experimental
ground-state binding energy \cite{audi95} of $^{4}$He  and the Coulomb
contributions  taken
from Ref. \cite{wolt90}, where they  were determined from a least-squares
fit to
experimental data.

Table I shows that all calculated binding energies decrease as $b_{\rm out}$
increases.
This is an obvious  consequence of the fact that most matrix elements of
$V_{\rm eff}$ become less attractive
when increasing  $b_{\rm out}$.
As regards the comparison with the experimental data, we see that for the
two lowest values of
$b_{\rm out}$ all
binding energies are significantly overestimated by our calculations.
A value of $b_{\rm out}=2.0$ fm brings the calculated binding energies for
Li isotopes and
their corresponding mirror nuclei into good agreement with experiment, the
discrepancies
ranging from 0.3 to 0.6 MeV. As regards the He isotopes (and their mirror
nuclei) a larger value of
$b_{\rm out}$ (2.25 fm) is needed to reproduce the experimental energies.
On the other hand, by increasing  $b_{\rm out}$ from 1.75 to 2.0 fm, the
calculated
binding energies of $^{8,9}$Be and $^{9}$B are shifted from 1-2 MeV
above to  4-5 MeV below the experimental values.
This indicates that the optimum value of $b_{\rm out}$ for these nuclei lies
between 1.75 and 2.0 fm. It turns out that it is 1.9 fm.

Note that in the above analysis we have
not tried to adjust the value of  $b_{\rm out}$ for each nucleus, but have
been
satisfied with discrepancies of a few hundred keV between experiment and
theory.
We would like to point out that the optimum
value of $b_{\rm out}$ is related to the nuclear binding energy
(relative to $^{4}$He) per valence nucleon. In fact, this quantity is almost
constant for nuclei which
require the same value of $b_{\rm out}$. More precisely, it is a
few hundred keV for the He isotopes, about 2-4 MeV for the Li isotopes, and
6-7  MeV for $^{8,9}$Be and $^{9}$B. The same situation occurs for all the
corresponding mirror nuclei.

Based on these findings,  we have found it appropriate to calculate the
spectra and
electromagnetic properties of the various  nuclei reported in Table I by
using the
values of  $b_{\rm out}$ derived from the above analysis.
We have verified that use of values of $b_{\rm out}$
different from the adopted ones leads to
an  overall worse  agreement  between  experimental and calculated spectra.
However, states with $T > T_{z}$ require a separate discussion, which will
be given at
the end of this Section.

Here we focus attention on $^{6-8}$Li and their corresponding
mirror nuclei. In Figs. 3-5  we compare the experimental spectra
\cite{nndc}with the
calculated ones ($b_{\rm out}=2.0$ fm). While the observed spectra of
$^{7}$Li and $^{7}$Be are quite
similar (the only significant difference is the absence
of a second $\frac{7}{2}^-$ state in the latter one),
the experimental information for $^{8}$B is very scanty.
For this reason, the following discussion will only concern Li isotopes.

As a general remark, we see that in the considered energy regions our
calculations give rise to
all the observed levels for each of the three nuclei. However, while for
$^{6}$Li and $^{7}$Li no more levels than the observed ones are predicted by
the theory,
for $^{8}$Li we find several  states without an experimental counterpart.

Let us now make some more specific comments on each Li isotope separately.
The ground state of $^{6}$Li is stable while the first excited state with
$(J^{\pi};T)=(3^{+};0)$ is just above the threshold for breakup into $\alpha
+d$ and
has a narrow width of 24 keV. The other two $T=0$ states have, instead,
fairly large widths ($\Gamma > 1000 ~{\rm keV}$). The $0^+$ state at 3.6 MeV
is
the isobaric analog of the  ground state in $^{6}$He and in $^{6}$Be, while
the $2^+$ state at 5.4 MeV
is the analog of the first excited state.
>From Fig. 3 we see that the first excited state
is very well reproduced by the theory. As regards the other two $T=0$
states, our calculation
overestimates the experimental excitation energies by more than 1 MeV, while
the
$(0^{+};1)$ and $(2^{+};1)$ states are underestimated by about 1.2 and 0.3
MeV, respectively.

The spectrum of $^{7}$Li contains the stable ground state with
$(J^{\pi};T)=(\frac{3}{2}^{-};\frac{1}{2})$ and
the $(\frac{1}{2}^{-};\frac{1}{2})$
first excited state, which decays by $\gamma$ emission. All other excited
states lie above
the threshold for breakup into $\alpha +t$, but only the
$(\frac{3}{2}^{-};\frac{1}{2})$ at
9.8 MeV is a broad resonance with  $\Gamma \gg 1200 ~{\rm keV}$.
The $T=\frac{3}{2}$ state at 11.2 MeV with a width of 260 keV is the
$T_z{}=\frac{1}{2}$ member of an isobaric quartet.
The analog states  with $|T_{z}|=\frac{3}{2}$
are the ground states of $^{7}$He and  $^{7}$B, while the member with
$T_{z}=-\frac{1}{2}$
is the state at 11.0 MeV in  $^{7}$Be.
The quantitative agreement between calculated and experimental
excitation energies is very satisfactory for all the levels, the only
exceptions
being the second $(\frac{3}{2}^{-};\frac{1}{2})$ state and the
$(\frac{3}{2}^{-};\frac{3}{2})$
state. In fact, the discrepancies are about 1 and 3 MeV for the former and
the latter states,
respectively, while they are less than few a hundred keV
for all the other states.

Turning to $^{8}$Li, the ground and first excited state are very
stable against the breakup,
the former decaying by $\beta^{-}$ emission. The second
excited state lies just above the threshold for
breakup into $^{7}{\rm Li} +n$ and is
fairly narrow with a width of 33 keV. A number of higher excited states have
been identified, some of them with large widths. In particular, the
$(1^{+};1)$
state at 3.2 MeV excitation energy and the $((3);1)$ state
at 6.1 MeV have widths $\Gamma \gg 1000 ~{\rm keV}$. The $(0^{+};2)$
isobaric
analog of the $^{8}$He ground state occurs at 10.8 MeV with a width less
than 12 keV.
>From Fig. 5 we see that not only  the first four calculated levels are in
the right order but also the excitation energies are in
very good agreement with experiment. Above these levels and up to
6 MeV  our calculation predicts four states, three of them
without an experimental counterpart. More precisely, we have  two
$(2^{+};1)$ states and  two states with  $(J^{\pi};T)=(0^+;1)$ and
$(1^{+};1)$, respectively, while only one
experimental level with spin equal to 0 or 1 is available in this energy
region.
Our calculation  suggests that this state, which lies at 5.4 MeV, has
$J^{\pi}=1^+$.
Between 6 and 8 MeV three levels have been observed, and the same number is
predicted by our calculation. Among them only one has a firm spin-parity
assignment and can be safely identified with the calculated $(4^{+};1)$
state,
whose excitation energy is only 80 keV larger than the experimental value.
As regards the $((3);1)$ level and that at 7.1 MeV with unknown spin and
parity,
we propose the assignment  $(3^{+};1)$ and  $(1^{+};1)$, respectively. In
this case,
the excitation energy of the latter state is almost exactly reproduced while
that of the former one is overestimated by about 1 MeV. Finally, we see that
the calculated
$(0^{+};2)$ level lies about 3 MeV below the experimental one.

>From the above we can conclude that, as regards the binding and excitation
energies, the overall
agreement between theory and experiment may be considered
quite satisfactory. In fact, significant discrepancies occur only for the
excitation energies
of states with fairly large widths or with $T > \mid T_{z} \mid$. As regards
these
latter states some comments are in order. The  $(0^{+};1)$,
$(\frac{3}{2}^{-};\frac{3}{2})$,
and  $(0^{+};2)$ states in $^{6}$Li, $^{7}$Li, and $^{8}$Li, respectively,
are isobaric
analogs of the ground states of  $^{6}$He, $^{7}$He, and $^{8}$He.
The  $(2^{+};1)$ in  $^{6}$Li is a member of the isospin triplet which is
comprised of the first
excited state in  $^{6}$He and  in $^{6}$Be.
At the beginning of this Section, we have shown that for the He isotopes a
larger value of $b_{\rm out}$
is required as compared to that adopted for the Li isotopes. We have then
found it appropriate
to calculate the energies of the $T > \mid T_{z} \mid$ states
in Li isotopes by making use of  $b_{\rm out}=2.25$ fm. It has turned out
that
all the new calculated excitation energies (relative to the ground-state
energies obtained with
$b_{\rm out}=2.0$ fm) go in the right direction largely reducing the
discrepancies with the experimental data.

Let us now come to the electromagnetic observables. In Table II
the measured moments \cite{stone} together with the $E2$
and $M1$ transition rates \cite{nndc} for $^{6-8}$Li and $^{8}$B
are compared with the calculated values. In our calculations
no effective charge has been attributed to the proton and neutron, and use
has
been made of free gyromagnetic factors.
 We have also calculated electric and magnetic effective
operators including only diagrams first order in $G$ \cite{cora99}.
We have found that the results do not
significantly differ from those obtained with bare operators.
This is not surprising, as our effective operators
take essentially into account the core-polaritation effects, which,
as pointed out in Sec. II,
are largely suppressed for  $b_{\rm out}$ significantly
larger than $b_{\rm in}$.

From Table II we see that the experimental magnetic moments and the $B(M1)$
values are very
well reproduced by our calculations. As regards the electric observables,
the agreement is not of
the same quality. However, while  our calculations underestimate the $E2$
transition rates as well as the quadrupole
moments, they reproduce the signs of the latter quantities (the sign
of the the quadrupole moment of $^{8}$B has not been measured).

\section{Summary}

In this paper, we have described how to calculate, for a chosen free
nucleon-nucleon potential,
the Brueckner $G$ matrix in a space composed of harmonic oscillator wave
functions
of two different length parameters $b_{\rm in}$ and $b_{\rm out}$, one for
the inner core orbits
and the other for the outer valence orbits.
Using this $G$ matrix the model-space effective interaction $V_{\rm eff}$ is
then derived
within the framework of the folded-diagram method.  Starting from the Bonn-A
potential
we have constructed an effective interaction for the $0p$ shell with a $G$
matrix corresponding to
the space specified by $b_{\rm in}=1.45$ fm  for the $0s$ core orbit and a
longer length parameter $b_{\rm out}$ for all the valence orbits (see Sec.
II).  The second-order core polarization contribution to the effective
interaction turns out to be  largely suppressed when
$b_{\rm out}$ is sufficently larger than $b_{\rm in}$. We have
also calculated some
typical third-order diagrams and we have found that, in this situation,
they are  very small. This shows  that
the effective interaction can be derived
in a very accurate way using the first- and second-order $G$-matrix
diagrams.
Similar suppression of core polarization effects was also observed
in our TFSM calculation of electromagnetic observables.

By employing this effective interaction we have performed a
shell-model study of
nuclei with $6 \leq A \leq 9$. To start with, we have analyzed the
dependence of the ground-state
binding energies on the value of $b_{\rm out}$. It turned out that the
binding energies for all the considered nuclei can be quite satisfactorily
reproduced by using three values
of $b_{\rm out}$. In particular, we have found that nuclei having  about
the same nuclear binding energy (relative to $^{4}$He) per nucleon require
the same value of $b_{\rm out}$.
We have then focused attention on the spectra of Li isotopes and their
mirror nuclei, which were calculated
by using $b_{\rm out}=2.0$ fm.
A good overall agreement between theory and experiment  is obtained,
significant  discrepancies existing
only for the energies of resonant states with fairly large widths and for
states with $T > T_{z}$. As regards the latter,
we have shown that they can be better described by making use of a larger
value of $b_{\rm out}$ (see discussion in Sec. III).
Finally, the electromagnetic observables, calculated using bare operators,
were compared with experiment. While
the dipole moments and the  $M1$ transition rates are in remarkably good
agreement with the measured values, the
experimental electric observables are all underestimated
by our calculations. Note that the theoretical values may
be brought into  agreement with experiment by using an effective proton
charge $e_{p}^{\rm eff}=1.5e$.

To conclude, we have shown that most properties of the $p$-shell nuclei can
be satisfactorily explained  making use
of a realistic effective interaction within the framework of the TFSM.  As
already mentioned in the Introduction,
several $0 \hbar \omega$ shell-model calculations have been performed for
these nuclei since the mid  1960s,
the most popular one being that of Cohen and Kurath \cite{cohen65}.
For all the  nuclei considered in the present paper,
the agreement with experiment is overall better than that
obtained in Ref. \cite{cohen65}. More gratifying, however, is the fact that
our study yields results which are comparable to, and in same cases even
better than,  those obtained from large-basis shell-model calculations. In
fact, on the one hand
we have obtained an agreement  with experiment which is
quite similar to that of Ref. \cite{wolt90b}, where a complete $(0+2) \hbar
\omega$ and an empirical effective interaction
were used. On the other hand, our calculations give a more satisfactory
description
of the $p$-shell nuclei than  that provided by
the large-basis no-core shell-model calculations of Ref.  \cite{nav98},
which  make use of effective
interactions derived from a modern $NN$ potential.
This indicates that in the TFSM approach  most of the effects which are
not explicitly taken into account in the  model space are included in the
effective interaction.

\acknowledgments

This work was supported in part by the Italian Ministero dell'Universit\`a e
della Ricerca Scientifica e
Tecnologica (MURST) and by the U.S. DOE Grant No. DE-FG02-88ER40388. NI
thanks the
European Social Fund for financial support.

\begin{figure}
\caption{First- and second-order $\hat Q$-box diagrams.}
\end{figure}

\begin{figure}
\caption{Dependence of the second-order core-polarization diagram $G_{3p1h}$
on $b_{\rm out}$.}
\end{figure}

\begin{figure}
\caption{Experimental and calculated levels of $^{6}$Li.}
\end{figure}

\begin{figure}
\caption{Experimental and calculated levels of $^{7}$Li and $^{7}$Be.}
\end{figure}

\begin{figure}
\caption{Experimental and calculated levels of $^{8}$Li and $^{8}$B.}
\end{figure}

\twocolumn[
\hsize\textwidth\columnwidth\hsize\csname @twocolumnfalse\endcsname
 
\mediumtext
\begin{table}
\setdec 0.00
\caption{Experimental and calculated ground-state binding energies (MeV).
See text for details.}
\begin{tabular}{ccccccc}
{   {$^{A}Z$} }   &
{   { Expt} } &
{   {TFSM(1.45)}} &
{   {TFSM(1.75)}} &
{   {TFSM(2.00)}} &
{  {TFSM(2.25)}} &
{   {TFSM(2.50)}} \\
\tableline

$^{6}$He & 29.27 & 32.84  & 31.53 & 30.58 & 29.76 &  29.13\\

$^{6}$Be & 26.92 & 30.48 & 29.17 & 28.22 & 27.40 &  26.67 \\


$^{6}$Li & 31.99 & 35.93 & 33.70 & 32.25 & 30.97 & 29.93  \\


$^{7}$He & 28.82 & 33.04 & 31.20 & 30.16 &  29.26 & 28.56 \\

$^{7}$B & 24.72 & 28.78 & 26.94 & 25.90 & 25.00 & 24.30 \\


$^{7}$Li & 39.24 & 46.57 & 41.78 & 38.65 & 35.95 & 33.83 \\

$^{7}$Be & 37.60 & 44.91 & 40.12 & 36.99 & 34.29 &  32.17 \\


$^{8}$He & 31.41 & 38.85 & 35.51 & 33.53 &31.79 &  30.43 \\

$^{8}$C & 24.78 & 31.89 & 28.55 & 26.57 & 24.83 &  23.47 \\


$^{8}$Li & 41.28 & 51.04 & 44.80 & 40.96 & 37.71 & 35.16 \\

$^{8}$B &37.74 & 47.48 & 41.24 &37.40& 34.15 & 31.60\\


$^{8}$Be & 56.50 & 67.41 & 57.62 & 51.17 & 45.57 & 41.13 \\


$^{9}$He & 30.26 & 37.31 &34.20 & 32.13 & 30.03 & 28.26 \\

$^{9}$Li & 45.34 & 57.57 & 49.97 & 45.27 & 41.15 & 37.89 \\

$^{9}$C & 39.03 & 51.31 & 43.71 & 39.01 & 34.89 & 31.63 \\


$^{9}$Be & 58.16 & 72.39 & 60.98 & 53.88 & 46.73 & 42.88 \\

$^{9}$B & 56.31 & 70.40 & 59.08 & 51.98 & 44.83 & 40.98 \\

\end{tabular}
\end{table}

\mediumtext
\begin{table}
\setdec 0.00
\caption{Experimental and calculated $B(E2)$ and $B(M1)$ values
(W.u.), $Q$ moments ($e$mb), and $\mu$ moments (nm) in $^{6-8}$Li and
$^{8}$B.}
\begin{tabular}{cccc}
{   {Nucleus} }   &
{   {Quantity} } &
{   {TFSM}} &
{   {Expt.}} \\
\tableline

$^{6}$Li & $B(E2; 3^{+}_{1} \rightarrow 1^{+}_{1})$ & 4.8 & $16.5\pm 1.3$
\\

& $B(E2; 2^{+}_{1} \rightarrow 1^{+}_{1})$ & 4.5 & $6.8\pm3.5$ \\

& $Q(1^{+}_{1})$ &-0.60& $-0.83\pm0.08$  \\

& $\mu(1^{+}_{1})$ &+0.87& $+0.82\pm0.00$  \\

$^{7}$Li & $B(E2; \frac{1}{2}^{-}_{1} \rightarrow \frac{3}{2}^{-}_{1})$ &
5.9 &$19.7\pm 1.2$  \\

& $B(E2; \frac{7}{2}^{-}_{1} \rightarrow \frac{3}{2}^{-}_{1})$ & 2.5 &
$4.3$ \\

& $B(M1; \frac{1}{2}^{-}_{1} \rightarrow \frac{3}{2}^{-}_{1})$ & 2.50 &
$2.75 \pm 0.14$ \\

& $Q(\frac{3}{2}^{-}_{1})$ & -24.4 &$-40.0\pm0.3$  \\

& $\mu(\frac{3}{2}^{-}_{1})$ & +3.81 &$+3.26\pm0.00$  \\

$^{8}$Li & $B(M1; 1^{+}_{1} \rightarrow 2^{+}_{1})$ & 2.7 &$2.8\pm 0.9$\\

& $B(M1; 3^{+}_{1} \rightarrow 2^{+}_{1})$ & 0.21 &$0.29\pm0.13$\\

& $Q(2^{+}_{1})$ & +24 &$+32.7\pm0.6$  \\

& $\mu(2^{+}_{1})$ & +1.52 &$+1.65\pm0.00$  \\

$^{8}$B & $Q(2^{+}_{1})$ & +44 & $64.6\pm1.5$  \\

& $\mu(2^{+}_{1})$ & +1.15 &$+1.04\pm0.00$  \\

\end{tabular}
\end{table}

]

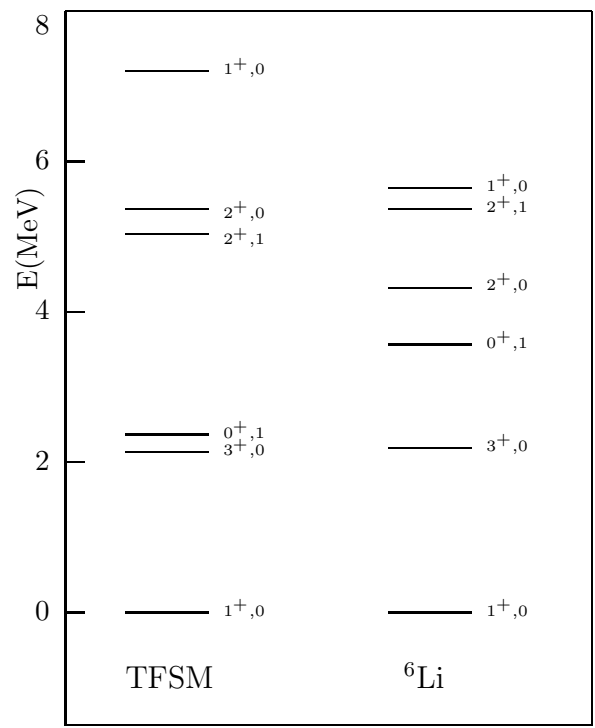
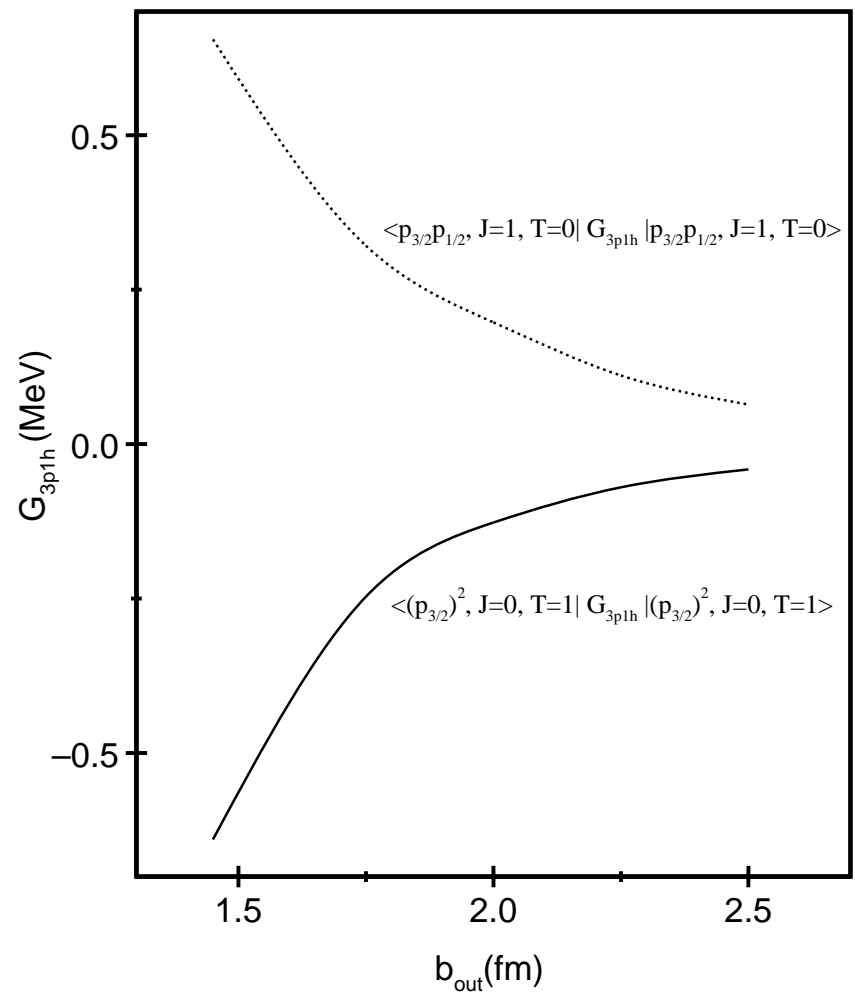
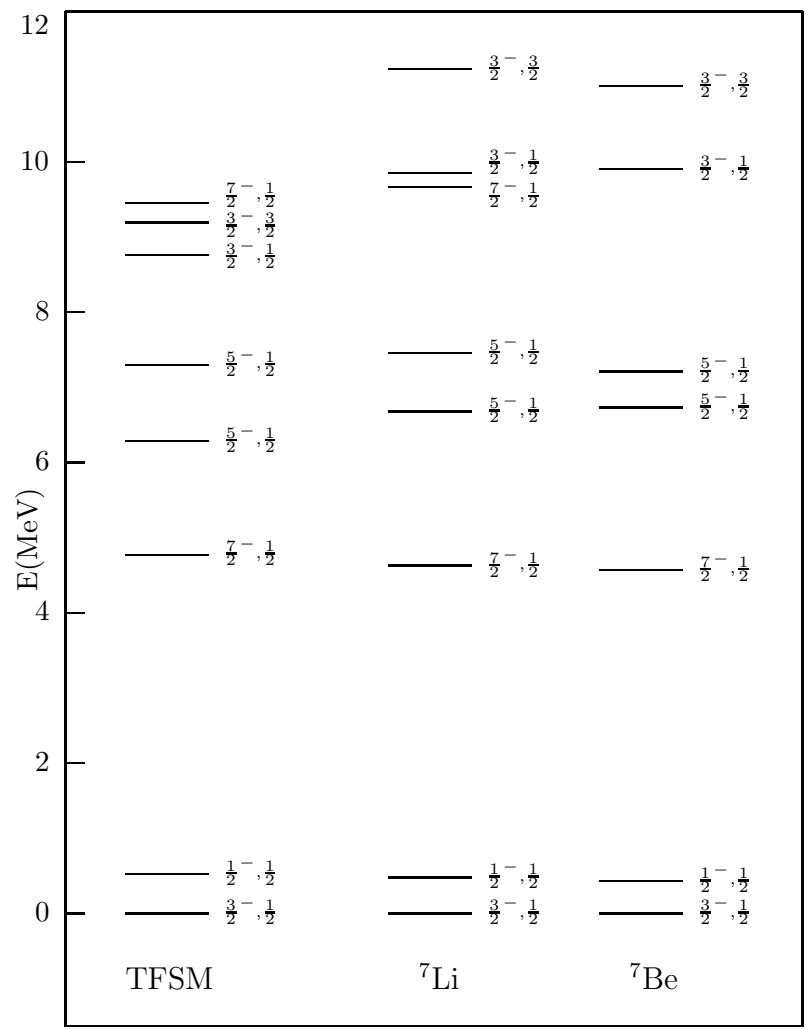
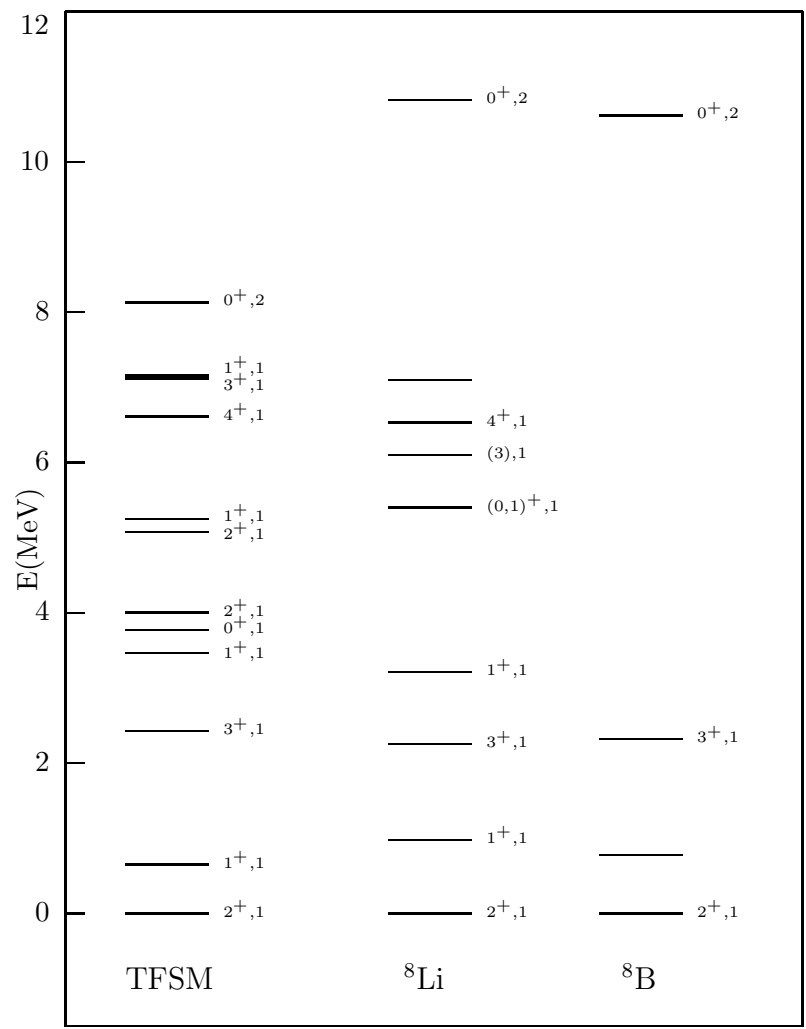
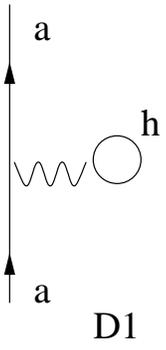
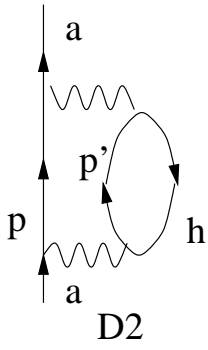
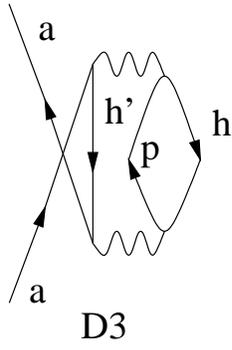
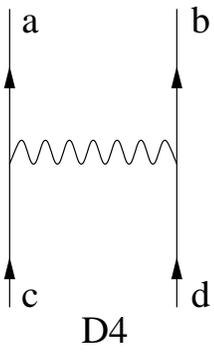
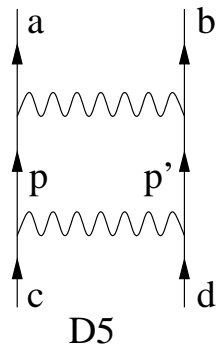
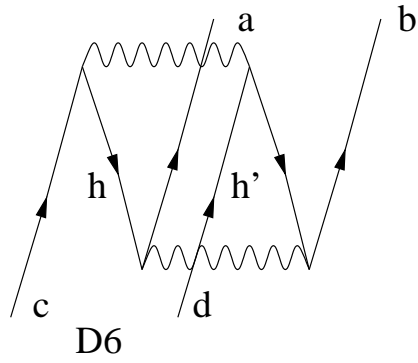
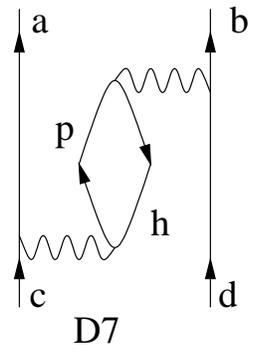
\end{document}